# Illusion of Control in a Brownian Game

J.B. Satinover[1] and D. Sornette[2]
[1]Laboratoire de Physique de la Matière Condensée
CNRS UMR6622 and Université des Sciences, Parc Valrose
06108 Nice Cedex 2, France
[2]Department of Management, Technology and Economics
ETH Zurich, CH-8032 Zurich, Switzerland
jsatinov@princeton.edu and dsornette@ethz.ch

**Abstract:** Both single-player Parrondo games (SPPG) and multi-player Parrondo games (MPPG) display the Parrondo Effect (PE) wherein two or more individually fair (or losing) games yield a net winning outcome if alternated periodically or randomly. (There is a more formal, less restrictive definition of the PE.) We illustrate that, when subject to an elementary optimization rule, the PG displays degraded rather than enhanced returns. Optimization provides only the illusion of control, when low-entropy strategies (i.e. which use more information) under-perform random strategies (with maximal entropy). This illusion is unfortunately widespread in many human attempts to manage or predict complex systems. For the PG, the illusion is especially striking in that the optimization rule reverses an already paradoxical-seeming positive gain—the Parrondo effect proper—and turns it negative. While this phenomenon has been previously demonstrated using somewhat artificial conditions in the MPPG (L. Dinis and J.M.R. Parrondo, Europhysics Letters **63**, 319 (2003); J. M. R. Parrondo, L. Dinis, J. Buceta, and K. Lindenberg, Advances in Condensed Matter and Statistical Mechanics, eds. E. Korutcheva and R. Cuerno, Nova Science Publishers, 2003), we demonstrate it in the natural setting of a history-dependent SPPG.

## A. Formalism of the Parrondo Effect (PE)

The Parrondo effect is the counterintuitive result where mixing two or more losing games can surprisingly produce a winning outcome. The basic Parrondo effect (PE) was first identified as the game-theoretic equivalent to directional drift of Brownian particles in a time-varying "ratchet"-shaped potential [1,2]. Consider $N > 1$ $s$-state Markov games $G_i$, $i \in \{1, 2, \ldots, N\}$, and their $N$ $s \times s$ transition matrices, $\hat{\mathbf{M}}^{(i)}$. For every $\hat{\mathbf{M}}^{(i)}$, denote the vector of $s$ conditional winning probabilities as $\vec{\mathbf{p}}^{(i)} = \{p_1^{(i)}, p_2^{(i)}, \ldots p_s^{(i)}\}$ and their steady-state probability vectors as $\vec{\Pi}^{(i)} = \left\{ \pi_1^{(i)}, \pi_2^{(i)}, \ldots, \pi_s^{(i)} \right\}$. For each game, the steady-state probability of winning is :

$$P_{win}^{(i)} = \vec{\mathbf{p}}^{(i)} \cdot \vec{\Pi}^{(i)} \quad (1)$$

Consider also a lengthy sequence of randomly alternating $G_i$ with individual time-averaged proportion of play $\gamma_i \in [0,1]$, $\sum_{i=1}^{N} \gamma_i = 1$. The transition matrix for the combined sequence of games is the convex linear combination $\hat{\mathbf{M}}^{(\gamma_1, \gamma_2, \ldots, \gamma_N)} \equiv \sum_{i=1}^{N} \gamma_i \hat{\mathbf{M}}^{(i)}$ with conditional winning probability vector $\vec{\mathbf{p}}^{(\gamma_1, \gamma_2, \ldots, \gamma_n)}$ and steady-state probability vector $\vec{\Pi}^{(\gamma_1, \gamma_2, \ldots, \gamma_n)}$. The steady-state probability of winning for the combined game is therefore

$$P_{win}^{(\gamma_1, \gamma_2, \ldots, \gamma_N)} = \vec{\mathbf{p}}^{(\gamma_1, \gamma_2, \ldots, \gamma_N)} \cdot \vec{\Pi}^{(\gamma_1, \gamma_2, \ldots, \gamma_N)} \quad (2)$$

A PE occurs whenever (and in general it is the case that) :

$$\sum_{i=1}^{N} \gamma_i P_{win}^{(i)} \neq P_{win}^{(\gamma_1, \gamma_2, \ldots, \gamma_N)} \quad (3)$$

i.e.,

$$\sum_{i=1}^{N} \gamma_i \vec{\mathbf{p}}^{(i)} \cdot \vec{\Pi}^{(i)} \neq \vec{\mathbf{p}}^{(\gamma_1, \gamma_2, \ldots, \gamma_N)} \cdot \vec{\Pi}^{(\gamma_1, \gamma_2, \ldots, \gamma_N)} \quad (4)$$

hence the PE, or "paradox", when the left hand side of (4) is less than zero and the right-hand side greater.

The original Parrondo game (PG) discussed in [1,2] employs three differently-biased coins. The coin to be tossed at a given time-step is determined by the net number of wins or losses. This "capital-dependent" PG can be expressed in terms of a $3 \times 3$ Markov transition matrix. The extension to PE games that are history-dependent was initially made in [3] with one history dependent game and the other not; Ref.[4] extends the concept to the linear convex combination of two history dependent games. A history-dependent game is one in which the prior *sequence* of wins and losses (of defined length *m*) determines which coin to toss, rather than the value of the accumulated capital. Similar to the use of the finite time-horizon τ in the time-horizon Minority Game (THMG) in [3, 4], a two-state memory-dependent process with memory of two bits is recast as a $2^2 = 4$-state memory-independent Markov process.

The most basic conditions under which a non-trivial, history-dependent PE might be sought are:
- $m = 2$
- $N = 2$
- $\gamma_1 = \gamma_2 = \frac{1}{2}$

- $P_{win}^{(1)} = P_{win}^{(2)} = \frac{1}{2} < P_{win}^{(1,2)}$
- $P_{win}^{(1)} = \vec{p}^{(1)} \cdot \vec{\Pi}^{(1)}$ and $P_{win}^{(2)} = \vec{p}^{(2)} \cdot \vec{\Pi}^{(2)}$

The value $m = 2$ means that the system has $2^2 = 4$ states: $\{00, 01, 10, 11\} \equiv \{1, 2, 3, 4\}$, with 0 = lose, 1 = win. The 4 steady state probabilities are simple algebraic functions of the conditional probabilities, up to a normalization factor:

$$\vec{\Pi}^{(1)} \propto \begin{pmatrix} (1-p_3^{(1)})(1-p_4^{(1)}) \\ (1-p_4^{(1)})p_1^{(1)} \\ (1-p_4^{(1)})p_1^{(1)} \\ p_1^{(1)} p_2^{(1)} \end{pmatrix}, \quad \vec{\Pi}^{(2)} \propto \begin{pmatrix} (1-p_3^{(2)})(1-p_4^{(2)}) \\ (1-p_4^{(2)})p_1^{(2)} \\ (1-p_4^{(2)})p_1^{(2)} \\ p_1^{(2)} p_2^{(2)} \end{pmatrix} \quad (5)$$

Here we have two fair games alternating at random in equal proportion to yield a winning game. Note that $\left(P_{win}^{(1,2)} - P_{win}^{(1)}\right) = \left(P_{win}^{(1,2)} - P_{win}^{(2)}\right) \equiv \Delta$, a finite amount. We suppose that it is possible to reduce all conditional probabilities in both games by some sufficiently small bias $\varepsilon$, such that:

- $P_{win}^{(1\varepsilon)}, P_{win}^{(2\varepsilon)} < \frac{1}{2} < P_{win}^{(1\varepsilon, 2\varepsilon)} < P_{win}^{(1,2)}$
- $P_{win}^{(1\varepsilon, 2\varepsilon)} - \frac{1}{2}\left(P_{win}^{(1\varepsilon)} + P_{win}^{(1\varepsilon)}\right) \equiv \Delta - \delta, \ 0 < \delta < \Delta$

We thus obtain two losing games alternating at random in equal proportion to yield a (still) winning game. We define two fair transition matrices $\hat{M}^{(1)}, \hat{M}^{(2)}$ (games) and bias matrix $\mathring{a}$ with $\gamma = \frac{1}{2}$:

$$\hat{M}^{(i)} = \begin{pmatrix} 1-p_1^{(i)} & 0 & 1-p_3^{(i)} & 0 \\ p_1^{(i)} & 0 & p_3^{(i)} & 0 \\ 0 & 1-p_2^{(i)} & 0 & 1-p_3^{(i)} \\ 0 & p_2^{(i)} & 0 & p_3^{(i)} \end{pmatrix} + \mathring{a}; \quad \mathring{a} = \begin{pmatrix} -\varepsilon & 0 & -\varepsilon & 0 \\ +\varepsilon & 0 & +\varepsilon & 0 \\ 0 & -\varepsilon & 0 & -\varepsilon \\ 0 & +\varepsilon & 0 & +\varepsilon \end{pmatrix} \quad (6)$$

Then

$$\hat{M}^{(\gamma, 1-\gamma)} = \gamma \hat{M}^{(1)} + (1-\gamma)\hat{M}^{(2)} = \frac{1}{2}\left(\hat{M}^{(1)} + \hat{M}^{(2)}\right) \quad (7)$$

The example provided in [5], and further analyzed in [6], will be extended here as well:

$$\{p^{(1)}\} = \{\tfrac{9}{10}, \tfrac{1}{4}, \tfrac{1}{4}, \tfrac{7}{10}\}; \quad \{p^{(2)}\} = \{\tfrac{1}{2}, \tfrac{1}{2}, \tfrac{1}{2}, \tfrac{1}{2}\} \quad (8)$$

Letting $\varepsilon = 0.005$ and substituting (8) in (6), (5) and (2), we obtain (as in [4]):

$$P_{win}^{(1)} = 0.4945; \ P_{win}^{(2)} = 0.4950; \ \tfrac{1}{2}\left[P_{win}^{(1)} + P_{win}^{(2)}\right] = 0.4947 < P_{win}^{(\tfrac{1}{2}, \tfrac{1}{2})} = 0.5010 \quad (9)$$

which illustrates quantitatively the Parrondo effect $P_{win}^{(1)} > 0.5$.

## B. Time Horizon PG (THPG): Reversal of the PE under Optimization

Ref.[5.6] present a capital-dependent multi-player PG (MPPG): At (every) time-step *t*, a constant-size subset of all participants is randomly re-selected actually to play. All participants keep individual track of their own capital but do not alternate games independently based on it. Instead, this data is used to select which game (all) selected participants must use at *t*. Given the individual values of the capital at $t-1$, the known matrices of the two games and their linear convex combination, the chosen game is the one which has the most positive expected *aggregate* gain in capital, summed over all participants.

This rule may be thought of as a static optimization procedure—static in the sense that the "optimal" choice appears to be known in advance. It appears exactly quantifiable because of access to each player's individual history. Indeed, if the game is chosen at random, the change in wealth averaged over all participants is significantly positive. But when the "optimization" rule is employed, the gain becomes a loss significantly greater than that of either game alone. The intended "optimization" scheme actually reverses the positive (collective) PE. The reversal arises in this way: The "optimization" rule causes the system to spend much more time playing one of the games, and individually, any one game is losing. This collective phenomenon is of interest as an example the phenomenon of "illusion of control." Here, we defined "illusion of control" as situations when low-entropy strategies (i.e. which use more information) under-perform random strategies (with maximal entropy).

However, the study of Ref.[5,6] has certain "artificial" features in both design and outcome. For example, all active players are constrained to "choose" the same rule. Such a constraint removes the example from the domain of complex (and most real-world) systems. The same applies when, as shown in [7], the enforced game is that which appears to maximize the wealth of (voted for by) the largest number of players; and even when the games being played are history-dependent. Second, the reversal of the PE occurs simply because the (enforced) "choices" turn out to be driven largely toward a single game. (Some authors have nonetheless attempted to draw social policy lessons from such artificial collective situations [8].) Of greater interest is the phenomenon presented in [9]: Collective games are shown to undergo a "current reversal" for certain mixing probabilities $\gamma_i$.

The term "current reversal" highlights the value of examining optimization rules in the setting of a PE, especially a positive one: An optimization rule that leads to a loss "against the current" of a positive PE is an especially good illustration of the illusion of control.

We present a pointed illustration of the illusion: under the most natural kind of optimization rule, a "current reversal" (reversal of a positive PE) appears in single-player PG's. This provides the most natural illustration of the illusion of control in PGs, and a suitable counterpoint to the analogous phenomenon in the natively collective MG as discussed in [10].

Furthermore, the reversal of the single-player PE under "optimization" *is not* caused by a significant imbalance of the system with respect to one game or another. For the PE in general, algorithms that actually do maximize the positive PE can be easily generated as the consequence of the Markov analysis presented above [5, 11]. These algorithms derive from the transition matrices which are known. They do not include the self-evident-seeming optimization rule that we employ here, on analogy to that employed in the Minority Games: At time $t$, play whichever game has accumulated the most points (wealth) over a sliding window of $\tau$ prior time-steps from $t-1$ to $t-\tau$.

We compare a numerical simulation using the same example games as in (6) through (9) (refs [3,4]) to their corresponding analytical formulation. Games $\hat{\mathbf{M}}^{(1)}$ and $\hat{\mathbf{M}}^{(2)}$ have memory $m = 2$ prior time steps. We include a time-horizon of length $\tau$ located before the history $\mu(t)$ in memory. At the start, a random history of three bits initializes play and the first $\tau$ steps require a random choice of game. In the simplest instance we let $\tau = 1$. The binary sequence of subsequent choices is thus dependent on a sliding window of prior binary wins/losses of length $m+\tau = 3$—the first such window of which $\mu(t)$ is not a subset. (Otherwise the exact winning game is defined in advance and the "optimization" succeeds trivially.) Whichever game would have won on the previous step had it been played (regardless of whether it actually was) is chosen to be played next. The player's wealth is based on the sequence of games actually played. If both games actually would have yielded the same outcome, win or lose, one of the two is chosen at random instead.

By construction, the individual games $\hat{\mathbf{M}}^{(1)}$ and $\hat{\mathbf{M}}^{(2)}$ played individually are both losing; random alternation between them is winning (the PE effect (4)) The one-player two-game history-dependent PE in our example is as follows: $\hat{\mathbf{M}}^{(1)}$ and $\hat{\mathbf{M}}^{(2)}$ have respective winning probabilities $P_{win}^{1} = 0.494$ and $P_{win}^{2} = 0.495$. Alternated at random in equal proportion $(\gamma_1 = \gamma_2 = 0.5)$, $P_{win}^{\gamma_1=0.5,\gamma_2=0.5} = 0.501$. We now express the choose-best optimization rule in Markov form. Under this rule, the two $s \times s$ matrices $\hat{\mathbf{M}}^{(1)}$ and $\hat{\mathbf{M}}^{(2)}$ do not combine as a linear convex $s \times s$ matrix sum. Instead, the combined game is represented by an $(s+\tau) \times (s+\tau)$ matrix $\hat{\mathbf{Q}}^{(1,2)}$. The $2s$ conditional winning probabilities are now $q_j = \frac{1}{2}\left\{p_{\alpha(j)}^{(1)}\left[1+p_{\beta(j)}^{(1)}-p_{\beta(j)}^{(2)}\right]+p_{\alpha(j)}^{(2)}\left[1-p_{\beta(j)}^{(1)}+p_{\beta(j)}^{(2)}\right]\right\}$ with $j = 1,2,\ldots,2s$ and indices $\alpha(j) = Mod[j-1,4]+1$, $\beta[j] = \frac{1}{2}(j-Mod[j-1,2]+1)$. (Under the choose-worst rule $q_j = \frac{1}{2}\left\{p_{\alpha(j)}^{(1)}\left[1-p_{\beta(j)}^{(1)}+p_{\beta(j)}^{(2)}\right]+p_{\alpha(j)}^{(2)}\left[1+p_{\beta(j)}^{(1)}-p_{\beta(j)}^{(2)}\right]\right\}$).

If the previously winning game is selected, $P_{win}^{best(1,2)} = 0.496$, while if the previously losing one is, $P_{win}^{best(1,2)} = 0.507$. Unexpectedly, choosing the previously best-performing game yields losses only slightly less than either $\hat{\mathbf{M}}^{(1)}$ and $\hat{\mathbf{M}}^{(2)}$ individually: The PE is almost entirely eliminated. Choosing the previously worst-performing games yields gains that exceed the PE proper.

The steady state probabilities for a simulation over 50 runs and 200 steps for each of the eight different possible initial states are shown in **Table 1**. The $R^2$ between the frequency of states obtained numerically and analytically is 0.988 over 40,000 runs.

Table 1: Analytically predicted and numerically simulated frequencies of the eight 3-state binary histories for a single player of two history-dependent Parrondo games under the "choose previously best" optimization rule. $R^2$=0.988 over 40,000 runs.

| $\pi$ | Analytic | Numeric |
|---|---|---|
| 000 | 0.075 | 0.072 |
| 001 | 0.164 | 0.165 |
| 010 | 0.172 | 0.165 |
| 011 | 0.093 | 0.097 |
| 100 | 0.164 | 0.165 |
| 101 | 0.101 | 0.097 |
| 110 | 0.093 | 0.097 |
| 111 | 0.138 | 0.142 |

The mechanism for this illusion-of-control effect characterized by the reversing of the PE under optimization is not the same as under a similar optimization rule for the Minority Game [10], as there is no collective effect and thus no-crowding out of strategies or games. (Nor is it the same as for the Multi-Player PG) As seen from (4), the PE proper results from a distortion of the steady-state equilibrium distributions $\vec{\Pi}^{(1)}$ and $\vec{\Pi}^{(2)}$ into a vector $\vec{\Pi}^{(\gamma_1,\gamma_2)}$ (for the n=2 version) which is more co-linear to the conditional winning probability vector $\vec{p}^{(\gamma_1,\gamma_2)}$ than are either individual game (this is just a geometric restatement of the fact that the combined game is winning). One may say that the random alternation of the two games tends on average to align these two vectors under the action of the other game. Choosing the previously best performing game amounts to removing this combined effect, while choosing the previously worst performing game tends to intensify this effect.

Consider the following simple illustration from [12], with $\gamma = \frac{1}{2}$:

$$\hat{\mathbf{M}}^{(1)} = \begin{pmatrix} \frac{5}{6} & \frac{1}{2} \\ \frac{2}{6} & \frac{1}{2} \end{pmatrix}; \quad \hat{\mathbf{M}}^{(2)} = \begin{pmatrix} \frac{1}{2} & \frac{1}{6} \\ \frac{1}{2} & \frac{5}{6} \end{pmatrix}; \quad \hat{\mathbf{M}}^{(1,2)} = \frac{1}{2}\left(\hat{\mathbf{M}}^{(1)} + \hat{\mathbf{M}}^{(3)}\right) = \begin{pmatrix} \frac{2}{3} & \frac{1}{3} \\ \frac{1}{3} & \frac{2}{3} \end{pmatrix} \quad (10)$$

From (6), (5) and (2):

$$P_{win}^{(1)} = \tfrac{1}{4}; \quad P_{win}^{(2)} = \tfrac{3}{4}; \quad P_{win}^{(1,2)} = \tfrac{1}{2} \quad (11)$$

There is no PE since

$$P_{win}^{(1,2)} = \tfrac{1}{2}\left(P_{win}^{(1)} + P_{win}^{(2)}\right) \quad (12)$$

The long term gain (loss) associated with $\hat{\mathbf{M}}^{(i)}$ is proportional to $P_{win}^{(i)}$. This implies unit positive or negative reward per time-step. We may however associate arbitrary, differ-ently-valued gains and losses. Suppose that for the transitions (elements) of both $\hat{\mathbf{M}}^{(1)}$ and $\hat{\mathbf{M}}^{(2)}$, we associate instead the following reward matrix $\hat{\mathbf{R}}$ [12]:

$$\hat{\mathbf{R}} = \begin{pmatrix} -1 & 3 \\ 3 & -1 \end{pmatrix} \tag{13}$$

Then the time-averaged unit change in wealth associated with $\hat{\mathbf{M}}^{(1)}$, $\hat{\mathbf{M}}^{(2)}$ and $\hat{\mathbf{M}}^{(1,2)}$ are computed as:

$$\Delta G^{(i)} = \vec{\Pi}^{(i)} \cdot \left( \hat{\mathbf{R}} \circ \hat{\mathbf{M}}^{(i)^\top} \right) \{1,1\}; \quad \left\{ \Delta G^{(1)}, \Delta G^{(2)}, \Delta G^{(1,2)} \right\} = \left\{ 0, 0, \tfrac{1}{3} \right\} \tag{14}$$

Thus, two fair games combine to make a winning game, a PE. The Hadamard product matrices $\left( \hat{\mathbf{R}} \circ \hat{\mathbf{M}}^{(i)^\top} \right)$ are not Markovian in that their elements are not probabilities, but products of a probability and a reward value. One may tinker with either the probabilities or the rewards to increase, decrease, eliminate or reverse the direction of the PE. The same result is obtained for any set of identical product values regardless of whether it is the probability or the reward that is thought of as altered.

Further light is shed on this phenomenon by considering a fully deterministic variant, i.e., where the elements of all $\hat{\mathbf{M}}^{(i)} \in \{0,1\}$. Consider indeed the following set of $\hat{\mathbf{M}}^{(i)}$, which are simply (one possibility for ) equations (10) rounded :

$$\hat{\mathbf{M}}^{(1)}_{Det} = \begin{pmatrix} 1 & 1 \\ 0 & 0 \end{pmatrix}; \quad \hat{\mathbf{M}}^{(2)}_{Det} = \begin{pmatrix} 0 & 0 \\ 1 & 1 \end{pmatrix} \tag{15}$$

$\hat{\mathbf{M}}^{(1)}$ and $\hat{\mathbf{M}}^{(2)}$ are now fully deterministic games. $\hat{\mathbf{M}}^{(1,2)}$ has the form of a single probabilistic game, but is indistinguishable from an alternation between games 1 and 2:

$$\hat{\mathbf{M}}^{(1,2)} = \begin{pmatrix} \tfrac{1}{2} & \tfrac{1}{2} \\ \tfrac{1}{2} & \tfrac{1}{2} \end{pmatrix} \tag{16}$$

The alternation may be periodic or equiprobably random. Furthermore, we may generate a strictly periodic sequence by imposing our "counteradaptive" optimization rule: Play at time $t$ the game that would have *lost* at time $t-1$. If $\hat{\mathbf{M}}^{(1)}$ and $\hat{\mathbf{M}}^{(2)}$ are multiplied by reward matrix $\hat{\mathbf{R}}$, the time-averaged unit changes in wealth associated with $\hat{\mathbf{M}}^{(1)}$, $\hat{\mathbf{M}}^{(2)}$ and $\hat{\mathbf{M}}^{(1,2)}$ are:

$$\Delta G^{(i)} = \vec{\Pi}^{(i)} \cdot \left( \hat{\mathbf{R}} \circ \hat{\mathbf{M}}^{(i)^\top} \right) \{1,1\}; \quad \left\{ \Delta G^{(1)}, \Delta G^{(2)}, \Delta G^{(1,2)} \right\} = \{-1, -1, +1\} \tag{17}$$

Thus, we have reproduced what looks like a (winning) PE by imposing a "paradoxical" optimization rule on the alternation of two wholly deterministic systems. As it happens for this example, the alternation may be itself a simple deterministic cycle. But the same results arise if the deterministic games are alternated at random with $\gamma = \tfrac{1}{2}$.

## C. Concluding remarks

In many social and economic activities, human agents attempt to maximize value. We often do so by adjusting our present strategy in accord with what has previously worked best. Yet this very adjustment often proves to have exactly the opposite effect—causing greater losses than if we had left well enough alone. A classic everyday example which has been analyzed in these terms is weaving in and out of traffic—we rarely gain, and often lose by doing so. We would do better sticking to whatever lane we find ourselves in [13]. The negative power of this effect is demonstrated by the perverse phenomenon which we have here highlighted as well: that in certain games, deliberately selecting what appears to be the worst approach can "paradoxically" enhance gains. While this effect follows directly in the MG (and in lane switching) from the "minority wins" rule, here a similar effect arises without requiring any such competitive mechanism.